\documentclass[aps, twocolumn, showpacs, amsmath, amssymb, reprint, prb,superscriptaddress]{revtex4}
\usepackage{graphicx}
\usepackage{bm}

\begin{document}

\title{Negative Refraction of Excitations in the Bose-Hubbard Model}
\author{R.A. Henry}
\affiliation{School of Physics, University of Melbourne, Parkville, Victoria 3010, Australia}
\author{J.Q. Quach}
\affiliation{School of Physics, University of Melbourne, Parkville, Victoria 3010, Australia}
\author{C.-H. Su}
\affiliation{Department of Infrastructure Engineering, University of Melbourne, Parkville, Victoria 3010, Australia}
\author{A.D. Greentree}
\affiliation{Applied Physics, School of Applied Science, RMIT University, Victoria 3001, Australia}
\author{A.M. Martin}
\affiliation{School of Physics, University of Melbourne, Parkville, Victoria 3010, Australia}

\date{\today}

\begin{abstract}
Ultracold atoms in optical lattices provide a unique opportunity to study Bose-Hubbard physics. In this work we show that by considering a spatially varying onsite interaction it is possible to manipulate the motion of excitations above the Mott phase in a Bose-Hubbard system. Specifically, we show that it is possible to ``engineer" regimes where excitations will negatively refract, facilitating the construction of a flat lens.
\end{abstract}


\maketitle

\section{Introduction}
The Bose-Hubbard system provides a useful theoretical and experimental platform to study the properties of quantum many-body systems and quantum phase transitions. One of the most dramatic implementations of the Bose-Hubbard model has been the prediction \cite{fisher1, jaksch1} and demonstration \cite{griener-02-qpt} of the Mott insulator to superfluid quantum phase transition in an ultracold atomic gas in an optical lattice. Such demonstrations are significant for applying canonical solid-state treatments to the more controllable regime of atom optics.

In this work, we study the phenomenon of negative refraction for excitations in the Bose-Hubbard system. Negative refraction of light can arise at the interface of negative index materials, in which the permittivity and permeability are engineered to be simultaneously negative \cite{veselago}. More generally, negative refraction arises when a wave moves between convex and concave surfaces in the bandstructure across an interface. Such an interface leads to all-angle negative refraction \cite{luo2002all}, which is not found in traditional media. Negative refraction has been demonstrated experimentally in the microwave regime \cite{smith1, shelby-01-evn, houck2003experimental, parazzoli2003experimental}. Through the application of transformational optics, band engineering of materials has become a rich platform for the control of electromagnetic waves, resulting in the realisation of an electromagnetic invisibility cloak \cite{schurig-06-mec}.
 
Recently, another class of metamaterials has become the focus of research: quantum metamaterials, in which global quantum coherence times exceed the signal transition time \cite{rakhmanov1}. In this field of research several schemes of qubit-array control have been proposed \cite{burgarth2011indirect,burgarth2009coupling,burgarth2010scalable,zagoskin2009quantum,zheludev2010road,felbacq2011commentary,hutter2011josephson,quach2011reconfigurable,zueco2012microwave,zagoskin2012superconducting}. The Bose-Hubbard system, realised as ultracold atoms in an optical lattice, provides a new example of a quantum metamaterial. Spatial variations in the onsite interaction strength provide the means of configuring the metamaterial, for excitations above the Mott phase ground state. In the case of ultracold atoms, an external magnetic field can be used to engineer the spatial dependence of the onsite interaction via Feshbach resonances. As discussed below, variations in the interaction strength locally change the bandstructure for quasi-particle excitations above a Mott phase ground state. Specifically, we demonstrate the existence of regimes where such variations cause quasi-particles to experience negative refraction.

The Bose-Hubbard Hamiltonian describes a system of interacting bosons on a lattice, and can be written as ($\hbar = 1$)
\begin{equation}\displaystyle H = -\kappa\sum_{\langle i,j \rangle}\hat{b}_j^\dagger \hat{b}_i + \dfrac{1}{2}\sum_iU_i\hat{n}_i (\hat{n}_i -1) + \mu\sum_i\hat{n}_i,\end{equation}
where the first summation is restricted to nearest neighbours,  $\hat{b}_i^{\dagger}$  ($\hat{b}_i$) is the bosonic creation (annihilation) operator, $\hat{n}_i=\hat{b}_i^{\dagger}\hat{b}_i$ is the occupation number operator, $\kappa$ is the hopping rate between sites, $\mu$ is the chemical potential and $U_i$ parameterizes the local strength of the  onsite interactions. Here we only consider a two-dimensional Bose-Hubbard system. For a uniform interaction strength ($U_i=U$) at temperature $T=0$, this Hamiltonian supports two quantum phases: the superfluid phase and the Mott insulator phase. In general, the ratio $\kappa /U$ determines the phase. The superfluid phase ($\kappa \gg U$) is characterised by the bosons being delocalised, forming a coherent wavefunction across the entire lattice. Conversely, the Mott phase ($U \gg \kappa$) is characterised by localisation and low coherence. The Mott phase also requires the mean number of bosons per site, referred to as the filling factor $g$, to be a positive integer. As this paper is exclusively concerned with the Mott phase, $g$ is always an integer that specifies which Mott state we are considering.

Here we study the scattering properties of small excitations above the Mott phase, specifically their scattering due to spatial variations in the on-site interaction $U$. The system considered is schematically shown in Fig. 1A: the lattice contains two regions $R_1$ and $R_2$, which differ only in that $U = U_1$ in $R_1$ and $U = U_2$ in $R_2$. In general an excitation starting in $R_1$ incident upon the boundary will scatter due to the abrupt change in $U$. By examining the bandstructures of the two regions, we show that negative refraction of the transmitted component of the excitation can be engineered across the interface. Reflection from the interface can be reduced by varying $U$ adiabatically, resembling the construction of a lens via a graded negative index \cite{quach2011reconfigurable}. We make use of this later in the sections concerning numerical simulations, however for now we will focus on the generic properties of refracted excitations, without considering reflection.

\begin{figure}
\centering
\includegraphics[width=8.5cm]{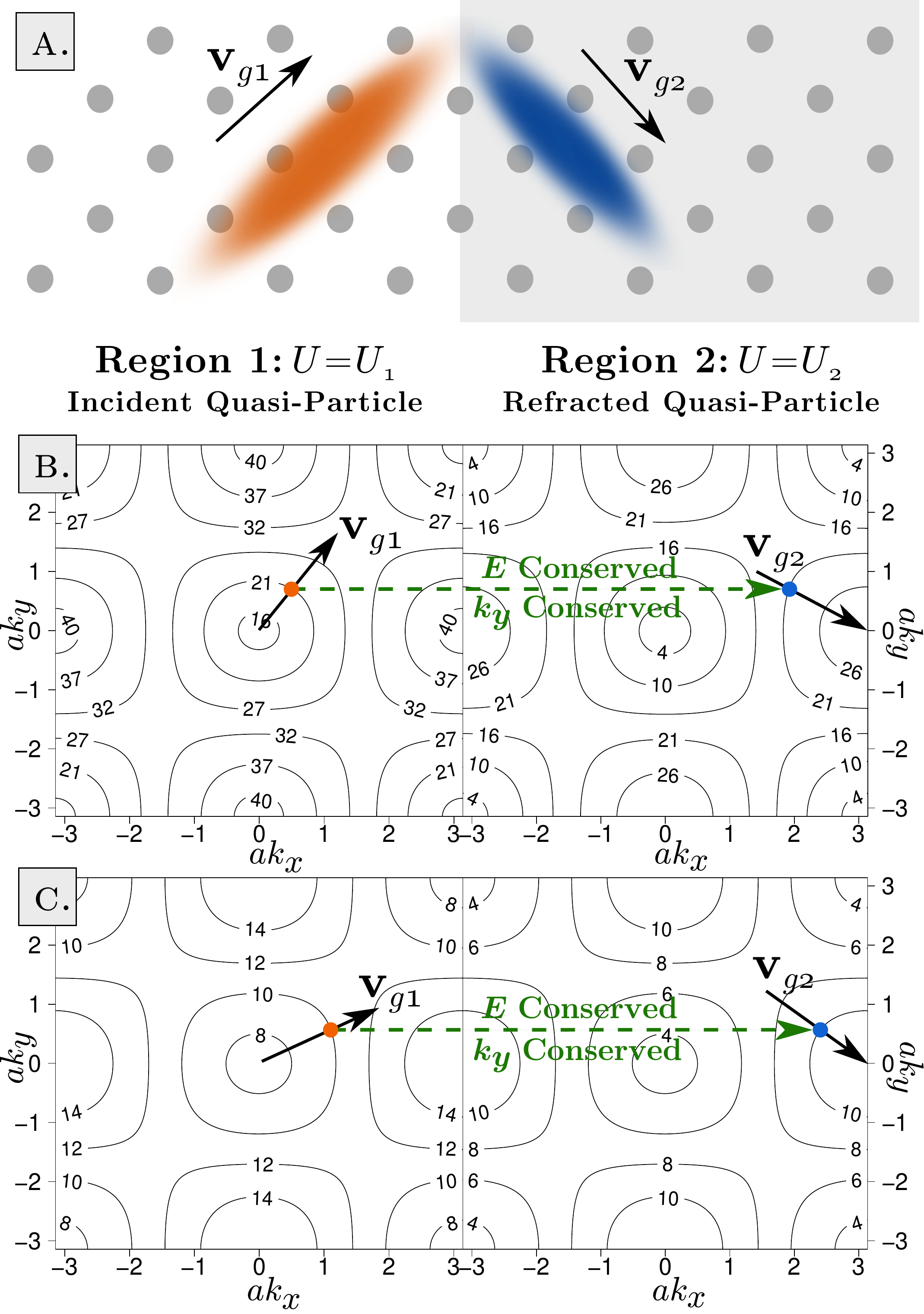}
\caption{(Color online) (A) Diagram of a rotated lattice separated into two regions with different interaction strengths $U_1$ and $U_2$. Imposed on this diagram is a schematic of quasi-particles refracting from the interface between the two regions. (B) Bose-Hubbard model bandstructures [Eq. (3)] of two regions, with $U_1=55\kappa$, $U_2=49\kappa$, $\mu=80\kappa$, $g=2$. A quasi-particle with energy $E=21\kappa$ moving across the interface between the two regions experiences negative refraction, as the $y$-component of the final group velocity is opposite in sign. (C) Single-excitation model bandstructures, also demonstrating negative refraction, with $\alpha_1=11.5\kappa, \alpha_2=7.5\kappa, g=1$. All energy contours, in B and C,  are in units of $\kappa$.} 
\end{figure}

\section{Excitations: Uniform Interaction}
Initially we consider the properties of excitations in a uniform Mott phase, i.e. $U$ is the same throughout the system. For a given integer filling factor $g$, the phase is determined by the values of $\mu$, $U$ and $\kappa$, so requiring the system to be in the Mott phase places a constraint on these parameters. The range of chemical potential values that give the Mott phase is bounded by $\mu_- < \mu <\mu_+$, where \cite{vanoosten-01-qpi}:
\begin{equation}
\mu_\pm=U\left(g-\dfrac12\right)-2\kappa \pm\sqrt{\dfrac{U^2}4-2U\kappa(2g+1)+4\kappa^2}.
\end{equation}
Assuming the condition given above is met, it is possible to consider propagating quasi-particle excitations above a background Mott state.  A quasi-particle (and equivalently, a quasi-hole) is a small-amplitude propagating distribution of fluctuations above and/or below the background state $|g\rangle \equiv \bigotimes_{i} |g_i\rangle \equiv \bigotimes_{\text{i}} (a_i^\dagger)^g |0\rangle$, where $i$ ranges over all sites. For a square lattice with uniform $U$, the dispersion relation for such quasi-particles and holes is \cite{vanoosten-01-qpi}
\begin{eqnarray}
E^{\pm} &=&-\mu + U\left(g- \dfrac{1}{2}\right) - \frac{\epsilon_{\bold{k}}}{2} \nonumber \\
&\pm&\frac{1}{2}\sqrt{\epsilon^2_\bold{k}-U\epsilon_\bold{k}(4g+2)+U^2},
\end{eqnarray}
where $\epsilon_{\mathbf{k}} = 4\kappa\,\textrm{cos}(k_xa)\,\textrm{cos}(k_ya)$ and $\epsilon_{\mathbf{k}} = 2\kappa[\textrm{cos}(k_xa)+\textrm{cos}(k_ya)]$ for the rotated and unrotated lattices respectively, with $a$ being the lattice spacing. $E^+$ gives the energy of a quasi-particle, while $E^-$ gives the energy of a quasi-hole. The (un)rotated lattice is defined by expressing the bosonic momenta in coordinates at (0)45 degrees to the natural coordinates of the square lattice. Experimentally, the particle number $N_\textrm{total}$ is fixed, which combined with the number of sites uniquely determines the chemical potential $\mu$. Therefore, the chemical potential $\mu$ is not an independent variable in Eq. (3). Nevertheless, we choose to keep our results in terms of the chemical potential, for simplicity.

Two additional experimental complications should be noted: firstly, the harmonic trapping potential typically used does not allow for the creation of a single uniform optical lattice, but rather results (in the 2D case) in a pattern of concentric circles of alternating superfluid and Mott phases \cite{bakr-10-pts}. Therefore, our results are applicable either within one of the Mott phase bands (with an additional external potential facilitating a change in $U$), or potentially between two adjacent bands, provided that the intermediate superfluid layer is sufficiently thin. In the future, realisations of effectively box-like trapping potentials may remove this issue. Secondly, we consider the scattering of single excitations. Therefore our results are in principle restricted to the regime where $k_BT << E^\pm$.

Below we consider the scattering of excitations through examination of the dispersion relation for quasi-particles. This can be considered both for the general case, given by Eq. (3), and in the limiting case of being deep in the Mott phase ($U >> \kappa$). In section III, we consider the general case, and then introduce, as required, the limiting case in section IV for use in simulations of negative refraction and lensing.

\section{Scattering of Excitations: Non-Uniform Interaction}  

In order to demonstrate the possiblity of negative refraction, we now focus on a system with an abrupt change in $U$ across some interface. Generally, negative refraction is the phenomenon of a wave being bent ``backwards", past the normal vector of the interface, as shown schematically in Fig. 1A. Equivalently, the component of the group velocity transverse to the interface changes in sign during refraction. The simplest way to establish negative refraction in our system is to examine the bandstructure directly. For a suitable choice of parameters, Fig. 1B plots the rotated lattice bandstructures for regions 1 and 2, according to Eq. (3). From these, negative refraction can be inferred as follows: consider a quasi-particle, in region 1, propagating towards the boundary ($\textbf{v}_{g1}$) with energy $E_1=21\kappa$. Upon scattering at the interface, the transmitted wave in region 2 has energy $E_2 = E_1 = 21\kappa$, due to conservation of energy. Continuity of the wavefunction (phase-matching), requires that the wavenumber transverse to the interface also be conserved. These two considerations, conservation of energy and phase-matching, are enough to uniquely determine the transmitted particle's place on the bandstrucutre. The group velocity $\textbf{v}_g = \nabla_\textbf{k}E$, perpendicular to the energy contours, has in this case changed sign in the $y$-direction, giving us negative refraction.

In general, it can be shown that quasi-particles and quasi-holes will negatively refract under a wide range of conditions, given an approriate change in $U$ between the two regions. Looking at one of the bandstructures in Fig. 1B, we immediately see distinct substructures composed of concentric, approximately circular contours, centered at each $\textbf{k}=(n\pi,m\pi); n, m \in \mathbb{Z}$. As shown in Fig. 1B, negative refraction occurs when the state jumps from the central substructure to an adjacent substructure, causing the necessary change in sign of $v_y$. Next, we see that the central substructure is separated from the adjacent ones by the contour at $k_x=\pm\frac\pi2$ and $k_y=\pm\frac\pi2$. Therefore, for a quasi-particle or quasi-hole to jump to a different substructure within the bandstructure, and hence negatively refract, the value of this contour must change. Specifically, it must change such that if it is initially greater than the quasi-particle or hole's energy, it must become less than that energy, and vice-versa. This boundary contour's value can be computed directly from the dispersion, noting that the cosine terms and hence $\epsilon_\textbf{k}$ are always $0$:
\begin{equation}
E^\pm|_{\text{boundary contour}} = -\mu + U\left(g-\dfrac12\pm\dfrac12\right)
.\end{equation}
We see that this expression can take any value if we assume that $U$ can be changed arbitrarily, meaning that negative refraction can occur in any Bose-Hubbard system over a discontinuity in $U$, provided the change in $U$ can be made sufficiently large. This must also be conditioned on the system remaining in the Mott phase, as per Eq. (2). There is one exception, however. In the case of quasi-holes (i.e. $E^-$) with $g=1$, all terms containing $U$ cancel, so the value of the boundary contour cannot be changed by a discontinuity in $U$. Therefore we have in this case the opposite result: negative refraction is never possible over a change in $U$ for quasi-holes with $g=1$.

\subsection{Negative Refraction: Algebraic Derivation}

Negative refraction has been demonstrated by considering the generic properties of the bandstructure in the two regions. To mathematically characterise the regimes of negative refraction we consider 
a quasi-particle excitation, in region $1$, incident upon the interface with wave-vector ${\bf k_1} = (k_{1x}, k_{1y})$ and velocity ${\bf v_1}$. This excitation will couple to an allowed mode of the lattice, in region $2$, and propagate with ${\bf k_2} = (k_{2x}, k_{2y})$ and ${\bf v_2}$. The refraction angle is $\theta_R = \textrm{arctan}(\tan k_{1y} \cot k_{2x})$. For an incident quasi-particle of energy $E_1$ and transmitted quasi-particle with energy $E_2$, energy conservation ($E = E_1 = E_2$) and phase matching ($k_{1y} = k_{2y} = k_y$) require that $k_{2x}$ satisfies the condition:
\begin{eqnarray}
4\kappa\cos (k_{y}a)\cos (k_{2x}a)=&\nonumber\\
&\hspace{-3.8cm}\dfrac{1}{E+U_2+\mu}\left\{-E^2+g^2U_2^2+E\left[\left(2g-1\right)U_2-2\mu\right]\right. \nonumber \\
&\hspace{-1.2cm}-\left. \mu\left[U_2+\mu\right]+gU_2\left[U_2+2\mu\right]\right\}
,\phantom{\hspace{1cm}}
\end{eqnarray}
where we have set the chemical potential ($\mu=\mu_1=\mu_2$) and the Mott filling factor ($g=g_1=g_2$) in the two regions to be the same and
\begin{eqnarray}
E&=&-\mu+U_1\left(g-\frac{1}{2}\right)- \frac{\epsilon_{\bf k_1}}{2}\nonumber \\
&\pm&\frac{1}{2}\sqrt{\epsilon_{\bf k_1}^2 -U_1\epsilon_{{\bf k_1}}\left(4g+2\right)+U_1^2}. 
\end{eqnarray}
\begin{figure}
\includegraphics[width=8.5cm]{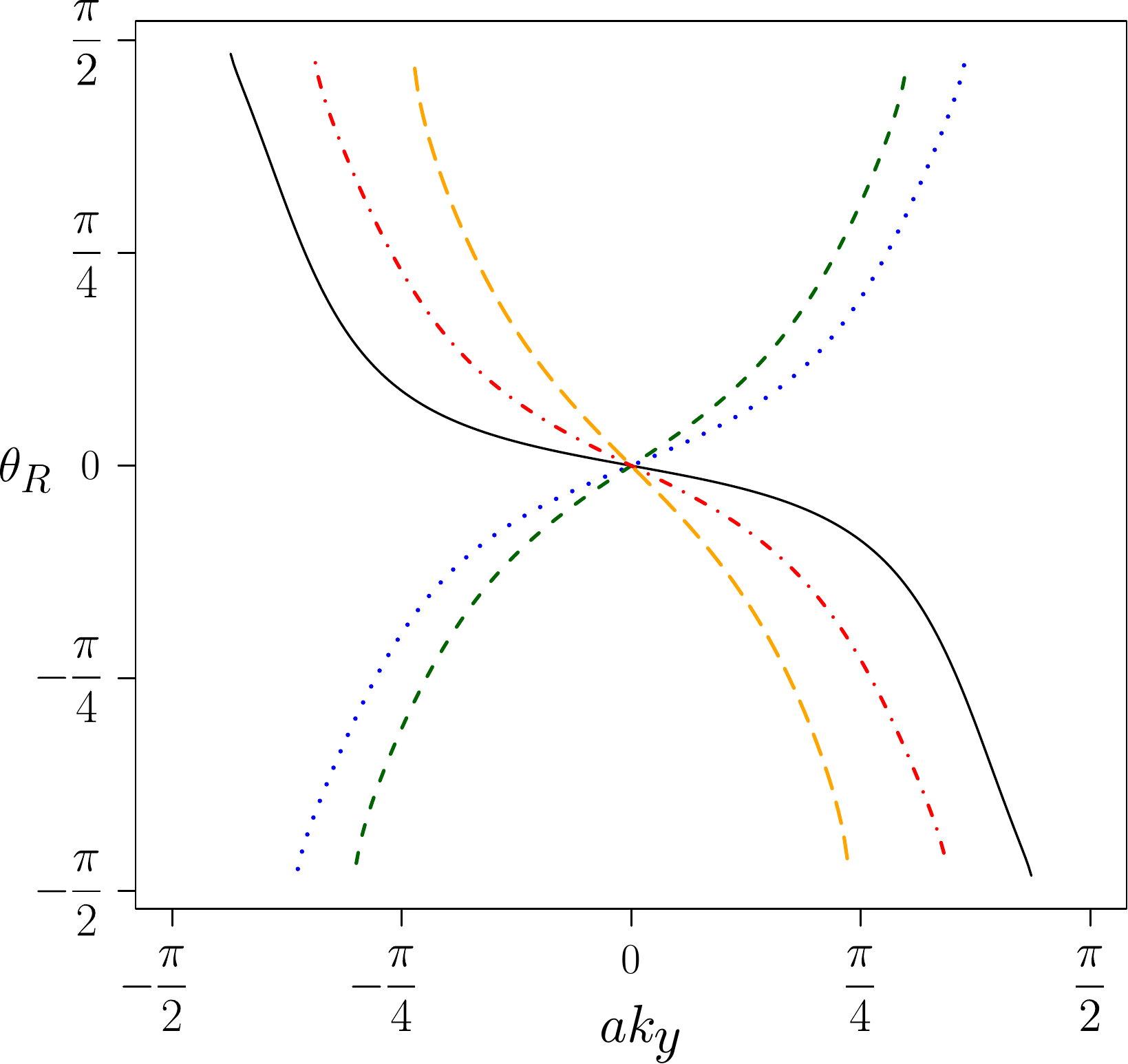}
\caption{(Color online) Angle of refraction (radians) versus transverse wavenumber for various sets of parameters, with $\mu=10.5\kappa$ in all cases. Negative refraction occurs when $ak_y$ is opposite in sign to $\theta_R$. Black solid line: quasi-particles with $U_2=25\kappa$, $g=1$, $E=16\kappa$. Red dotted-dashed line: quasi-particles with $U_2=11\kappa$, $g=2$, $E=16\kappa$. Blue dotted line: quasi-particles with $U_2=30\kappa$, $g=1$, $E=16\kappa$. Orange long-dashed line: quasi-holes, with $U_2=10\kappa$, $g=2$, $E=16\kappa$. Green short-dashed line: quasi-holes, with $U_2=24\kappa$, $g=1$, $E=-7.5\kappa$. The blue and green lines show ordinary refraction, while the others show negative refraction.} 
\end{figure}

For a given $E$, $U_2$, $\mu$ and $g$ it is now possible to compute the angle of refraction $\theta_R$. Figure 2 shows the dependence of $\theta_R$ on the transverse wavenumber of the incident wave, for various parameters. Specifically, we see that regimes of negative refraction are attainable for quasi-particles (solid black and red dashed-dotted curves) and quasi-holes (orange long-dashed curve). The results demonstrate all angle negative refraction, meaning that any incident quasi-particles of the specified energy $E$ will be negatively refracted at this potential discontinuity, regardless of its angle of incidence.

\subsection{Negative Refraction: Numerical Simulation}

Having established negative refraction analytically, we now provide a numerical demonstration. Since the full Bose-Hubbard model cannot be simulated except for small lattices, an approximation must be used. For simplicity, we will only consider quasi-particle, rather than quasi-hole, excitations. We assume that the quasi-particle excitation is entirely composed of small contributions to the $|(g+1)_i\rangle$ state at different sites, with $g$ being the filling factor of the background Mott state. This regime is equivalent to the deep Mott phase, where $U >> \kappa$. Therefore we are only modelling the dynamics of a lattice of two-level systems, the two states representing $|g_i\rangle$ and $|(g+1)_i\rangle$ of the Bose-Hubbard model (at particular sites). In this regime, a general state has the form $|\psi\rangle = \sum_r c_r |e_r\rangle$, where $|e_r\rangle$ is an excitation localised to a single site, given by $|e_r\rangle = |(g+1)_r\rangle\bigotimes_{i\neq r} |g_i\rangle .$ We refer to this regime as the single quasi-particle excitation manifold \cite{quach2009band}. The Hamiltonian is similar in form to the original Bose-Hubbard model, but the quadratic term becomes linear since there are only two states at each site:
\begin{equation}\displaystyle H_{TB} = -\kappa\sum_{i,j}\hat{b}_j^\dagger \hat{b}_i + \alpha\sum_i \hat{n}_i,\end{equation}
where $\hat{b}_i$ and $\hat{b}_i^\dagger$ are the creation and annihilation operators, which now only act on two-level systems representing $|g_i\rangle$ and $|(g+1)_i\rangle$ (at a particular site $i$), $\kappa$ is the hopping rate as before, and $\alpha$ is a constant combining the linear contributions of the chemical potential ($\mu$) and the on-site interaction strength ($U$) terms. Because we assume the $|(g+1)_i\rangle$ contributions that compose the quasi-particle are small in amplitude, $\alpha$ can be thought of as giving the first-order energy of this small $|(g+1)_i\rangle$ contribution. More precisely, $\alpha$ is the proportionality between the amplitude of the contribution and the change in the system's energy i.e. $\Delta E\approx \alpha\langle e_r|g\rangle$, where $|g\rangle$ is the background state. Therefore, $\alpha$ should be equal to the change in energy of a Mott phase with filling factor $g$ when we add a single boson to a single site: 
\begin{equation}
\alpha =-\mu+Ug.
\end{equation}
Note that changing $\alpha$ globally by some value only changes the system's total energy by an amount independent of the configuration, since particle number is conserved in the Hamiltonian. Thus a global change in $\alpha$ has no effect on the dynamics beyond an overall phase oscillation. However, we still still observe refraction if a wavepacket propagates between two regions of different $\alpha$. Equation (7) is identical to the Hamiltonian of the tight-binding model, with the following dispersion:
\begin{equation}
E^\pm = \dfrac{\alpha-\epsilon_{\textbf{k}}}2\mp\frac{\alpha+\epsilon_{\textbf{k}}}{2}.
\end{equation}

Figure 1C shows bandstructures for two regions in this model. By the same argument as before, we see that the single-excitation model will also exhibit negative refraction, as previously demonstrated by Su \cite{chsuthesis}. Thus, although it is a considerable simplification, the model is sufficient to capture the essential physics of negative refraction. 

In Fig. 3, we show negative refraction for an adiabatic change in $\alpha$ between two regions. The adiabaticity is demonstrated by the lack of reflection from the interface. The gradual change is created using $U(x)=\text{tanh}[10(x-x_0)/a]$, where $x$ is a coordinate perpendicular to the discontinuity. This is done to minimise reflection across the interface: a sharp change in $\alpha$ still produces negative refraction, but with a high degree of reflection. The initial state is a normalised Gaussian of the form $\exp{}(-b|\textbf{r}|^2-i\textbf{k}\cdot\textbf{r})$, with $b=0.008 / a^2$, $|\textbf{k}|=1.1/a$, and $\textbf{k}$ being at an angle of $\theta=0.485 $ radians to the $x$-axis, where $\textbf{r}$ is the displacement from the centre of the initial state. The factor of $\exp{}(-i\textbf{k}\cdot\textbf{r})$ imparts an initial momentum to the state. The arrows shown in Fig. 3 indicate the analytically determined directions of the initial and final group velocities. The direction of the initial group velocity is at $0.485$ radians to the $x$-axis i.e. the direction of the initially applied momentum. The final group velocity is determined using phase matching and energy conservation for the single-excitation model, as was done previously for the Bose-Hubbard system, which gives the result $\textbf{v}_g^f=(-v^i_{g\,x},v^i_{g\,y})$, giving us $\theta_{final} = -0.485$ radians in this case. This matches well with the final direction of propagation seen in the simulation.

\begin{figure}
\centering
\includegraphics[width=8.5cm]{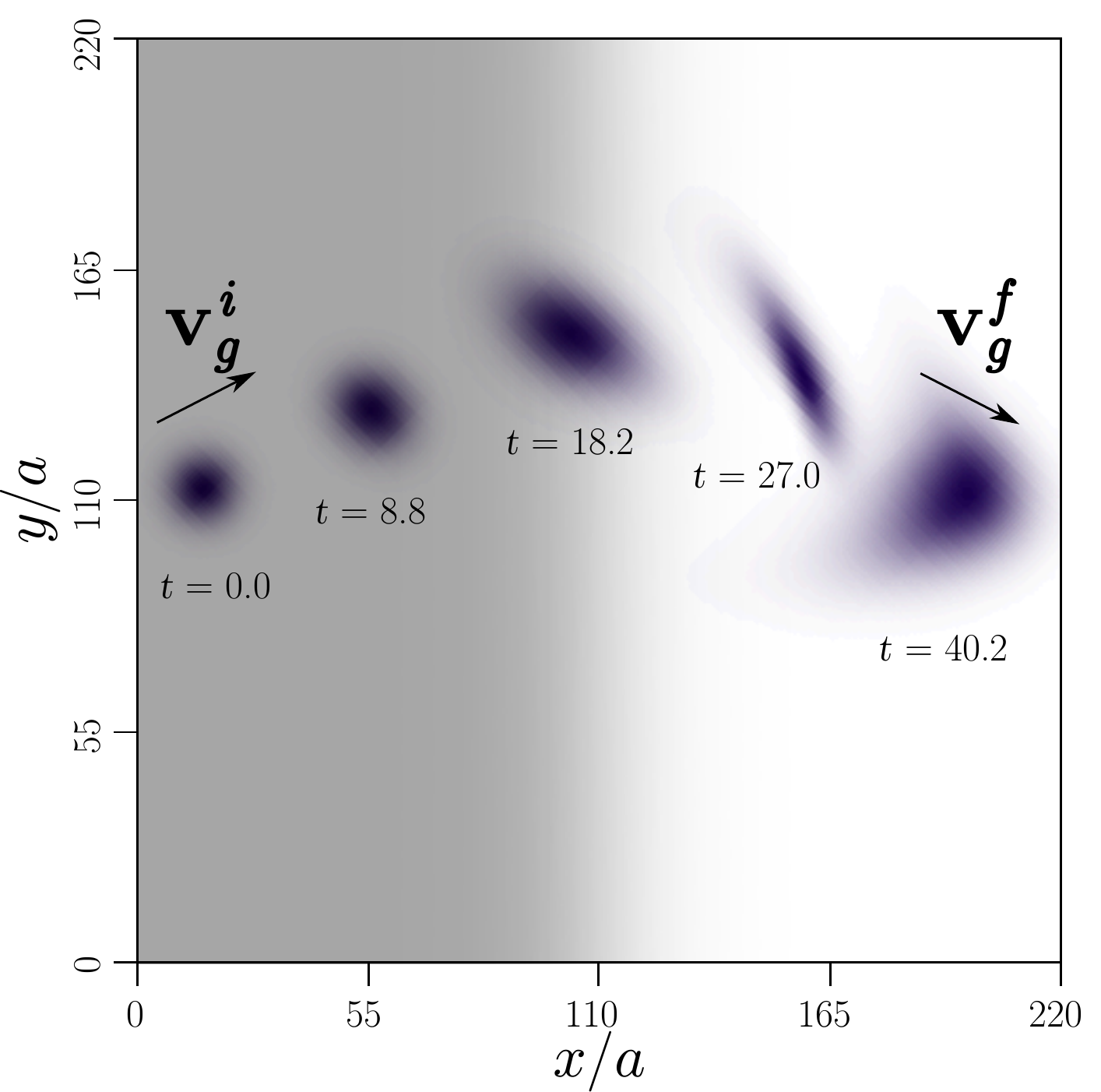}
\caption{(Color online) Results of a single-excitation model negative refraction simulation on a $200\times200$ lattice, for $\Delta\alpha = 7.0\kappa$. Five snapshots are shown, labelled with their simulation times. The shaded background shows $\alpha$, with white = 5.75$\kappa$ and grey = 12.75$\kappa$. The potential is varied gradually, as $\textrm{tanh}[10(x-x_0)/a]$, to reduce reflection. The arrows show the group velocity of the initial and final states. The initial velocity is shown as applied numerically, and the final velocity has been calculated analytically. The snapshots have been rescaled for clarity, with relative maximum intensitites of $1.00$,  $0.73$, $0.36$, $0.66$, and $0.22$, respectively. }
\end{figure}
\section{Lensing}
Having demonstrated negative refraction for excitations in Bose-Hubbard systems, we now consider the formation of flat lenses for such excitations. In optics, a flat lens can be constructed from a single band of negatively refractive material. We use a similar construction: a band-like region with a lower interaction strength $U$ focuses incident quasi-particles. This was demonstrated in a single-excitation model simulation as shown in Fig. 4. The initial state was a superposition of two states with opposite $k_y$ values, which were otherwise constructed in the same way as the initial state of the negative refraction simulation. This superposition creates the interference pattern that can be seen both in the initial state and in the final focused state. As in Fig. 3, $\alpha$ is varied smoothly to avoid significant reflection. The arrows in Fig. 4 indicate the directions of the group velocities initially, inside the lens, and after the lens. As before, the initial group velocity is at an angle of $\theta=0.485$ radians to the $x$-axis, and the velocities inside and after the lens are calculated as per $\textbf{v}_g^f=(-v^i_{g\,x},v^i_{g\,y})$.

\section{Conclusions}
Through two analytic methods, one based on bandstructure and one a direct derivation of the refraction angle, we have shown that negative refraction of quasi-particles and quasi-holes may occur at a discontinuity in the interaction strength $U$ in the Mott phase of the Bose-Hubbard system. Furthermore, we found that negative refraction is not possible for quasi-holes when the filling factor $g = 1$. We then demonstrated negative refraction numerically for a single-excitation model ($U >> \kappa$), which serves as an approximation to a small-amplitude quasi-particle moving across a stable background Mott state. Finally, we used negative refraction in the limit $U >> \kappa$ to numerically construct a flat lens under the single excitation model.

These results serve as a demonstration of the potential of the Bose-Hubbard system as a platform for a variety of novel quantum devices. For example, a series of lenses of the type demonstrated here could be used to guide or focus an atom laser \cite{robins} as it passes ``over" an optical lattice. The recent observation of quasi-particles of the kind described in this paper by \cite{bakr-10-pts} shows that such effects may soon be experimentally realisable. The required form of the interaction strength $U$ can in principle be produced via Feshbach resonances, using an external magnetic field.

\begin{figure}[h!]
\centering
\includegraphics[width=8.5cm]{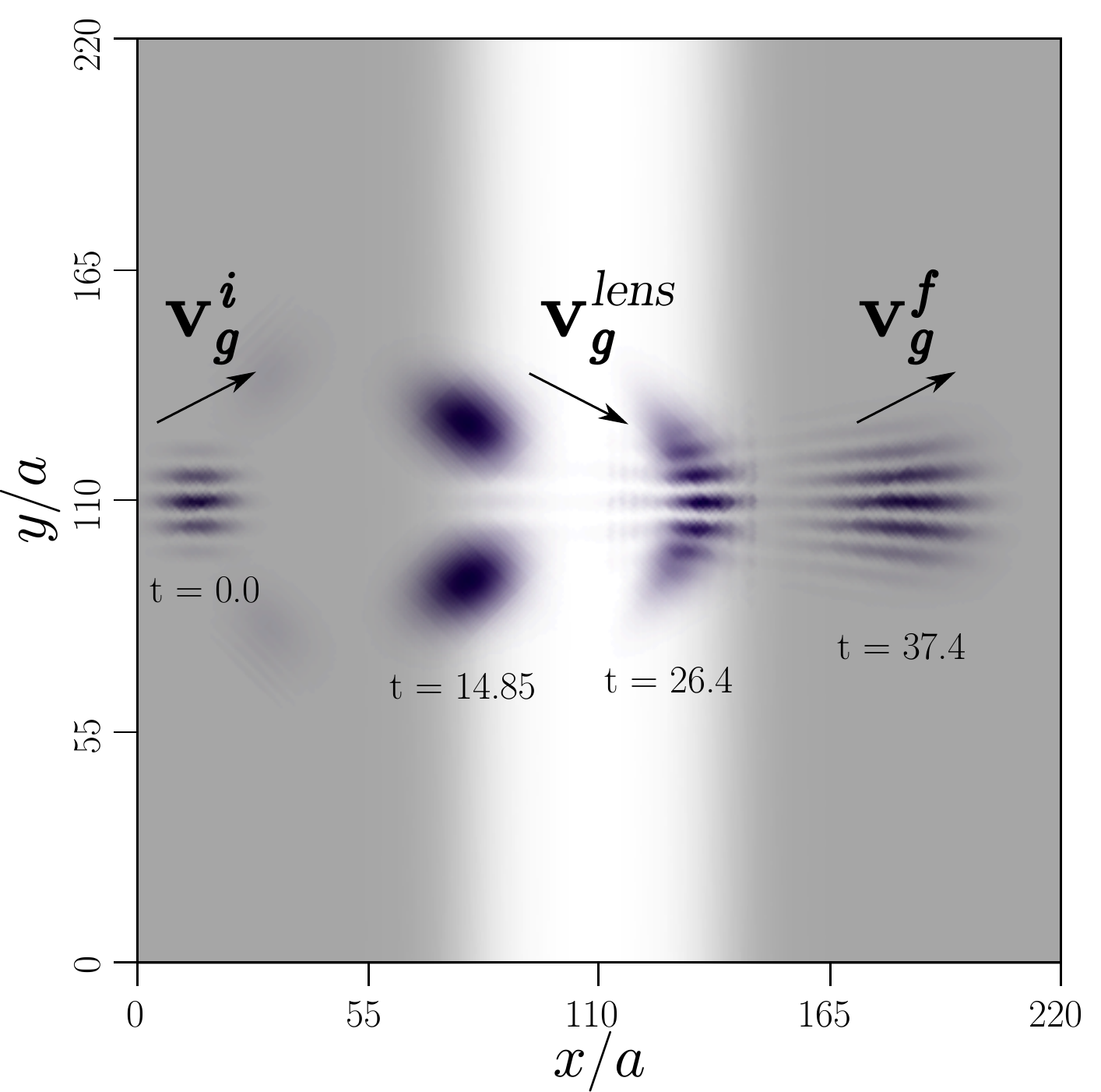}
\caption{(Color online) Results of a single-excitation model lensing simulation on a $200\times200$ lattice. Four snapshots are shown, labelled with their (arbitrary) simulation times. The background shows $\alpha$, with white = 5.75$\kappa$ and grey = 12.75$\kappa$. The arrows show the group velocites in the regions before, inside and after the lens. The snapshots have been rescaled for clarity, with relative maximum intensities of $1$, $0.152$, $0.291$ and $0.312$, respectively. }
\end{figure}

\section{Acknowledgements}
A.D.G. Acknowledges the Australian Research Council for financial support (project No. DP130104381).

\bibliography{paper}

\end{document}